\documentclass[10pt,aps,twocolumn,preprintnumbers,prd,noshowpacs,nofootinbib,noshowkeys,floatfix,superscriptaddress]{revtex4}
\usepackage[dvips]{graphics,graphicx}
\usepackage[colorlinks=true,linktocpage=true,linkcolor=blue,citecolor=blue]{hyperref}
\usepackage[usenames,dvipsnames]{color}
\usepackage{amsmath, amssymb,oldgerm}
\usepackage{multirow}
\usepackage{color}
\usepackage[normalem]{ulem}  
\usepackage{braket}
\usepackage{slashed}
\usepackage[mathscr]{eucal}

\newcommand{\n}{\nonumber}

\newcommand{\mC}{\mathfrak{C}}

\newcommand{\ms}{\mathfrak{s}}

\newcommand{\beq}{\begin{equation}}
\newcommand{\eeq}{\end{equation}}

\newcommand{\eqs}{Eqs.~}
\newcommand{\eq}{Eq.~}

\newcommand{\taum}{\tau}
\newcommand{\proj}{\Delta}

\newcommand{\mn}{\mathfrak{n}}

\newcommand{\map}{\mathfrak{p}}

\newcommand{\mft}{\mathfrak{p}}

\newcommand{\mfv}{\mathfrak{z}}
\newcommand{\mfw}{\mathfrak{q}}

\newcommand{\lmur}{{\langle\mu\rangle}}

\newcommand{\bbb}{\textcolor{black}}

\usepackage{cancel}
\usepackage{xcolor}
\usepackage{tikz}                   
\usepackage{pgfplots}				
\usepackage{caption}
\usepackage{subcaption}				

\definecolor{goethe-blau}{cmyk}{1.0,0.2,0.0,0.4}
\definecolor{hellgrau}{cmyk}{0.04,0.04,0.05,0.02}
\definecolor{sandgrau}{cmyk}{0.12,0.09,0.13,0.0}
\definecolor{dunkelgrau}{cmyk}{0.25,0.25,0.30,0.75}
\definecolor{emo-rot}{cmyk}{0.04,1.0,0.8,0.07}
\definecolor{purple}{cmyk}{0.08,1.0,0.3,0.36}
\definecolor{senfgelb}{cmyk}{0.01,0.25,1.0,0.05}
\definecolor{gruen}{cmyk}{0.62,0.4,0.87,0.09}
\definecolor{magenta}{cmyk}{0.08,0.86,0.12,0.12}
\definecolor{orange}{cmyk}{0.0,0.7,1.0,0.04}
\definecolor{sonnengelb}{cmyk}{0.0,0.12,0.95,0.0}
\definecolor{helles-gruen}{cmyk}{0.4,0.17,0.81,0.07}
\definecolor{lichtblau}{cmyk}{0.8,0.0,0.06,0.04}

\begin{document}

\title{Relativistic dissipative spin hydrodynamics from kinetic theory with a nonlocal collision term}
\author{Nora Weickgenannt}

\affiliation{Institute for Theoretical Physics, Goethe University,
Max-von-Laue-Str.\ 1, D-60438 Frankfurt am Main, Germany}

\author{David Wagner}

\affiliation{Institute for Theoretical Physics, Goethe University,
Max-von-Laue-Str.\ 1, D-60438 Frankfurt am Main, Germany}

\author{Enrico Speranza}

\affiliation{Illinois Center for Advanced Studies of the Universe and Department of Physics, University of Illinois at Urbana-Champaign, Urbana, IL 61801, USA}

\author{Dirk H.\ Rischke}

\affiliation{Institute for Theoretical Physics, Goethe University,
Max-von-Laue-Str.\ 1, D-60438 Frankfurt am Main, Germany}
\affiliation{Helmholtz Research Academy Hesse for FAIR, Campus Riedberg, Max-von-Laue-Str.\ 12, D-60438 Frankfurt am Main, Germany}

\begin{abstract}
We derive relativistic dissipative spin hydrodynamics 
from kinetic theory featuring a nonlocal collision term using the method of moments. 
In this framework, the components of the spin tensor are  
dynamical variables which obey relaxation-type equations. We find that the corresponding relaxation times are 
determined by the local part of the collision term, while the nonlocal part contributes to the Navier-Stokes terms 
in these equations of motion. The spin relaxation time scales 
are comparable to those of the usual dissipative currents. Finally, the Navier-Stokes limit of the Pauli-Lubanski 
vector receives contributions proportional to the shear tensor of the fluid, 
which implies that the polarization of hadrons observed in heavy-ion collisions is influenced by dissipative effects.
\end{abstract}

\maketitle

\textit{Introduction ---}
The Barnett effect, i.e., the polarization of a system through rotation, is well-known 
from condensed matter physics~\cite{Barnett:1935}. A similar effect has been 
proposed~\cite{Liang:2004ph,Voloshin:2004ha,Betz:2007kg,Becattini:2007sr} and subsequently 
observed~\cite{STAR:2017ckg,Adam:2018ivw} in the context of high-energy physics: hadrons, such as Lambda 
hyperons, are polarized by the nonvanishing orbital angular momentum in noncentral heavy-ion collisions. 
This phenomenon has spurred many new theoretical developments in the past years~\cite{Becattini:2013vja,Becattini:2013fla,Becattini:2015ska,Becattini:2016gvu,Karpenko:2016jyx,Pang:2016igs,Xie:2017upb,Becattini:2017gcx,Becattini:2020ngo,Florkowski:2019qdp,Florkowski:2019voj,Zhang:2019xya,Becattini:2019ntv,Xia:2019fjf,Wu:2019eyi,Sun:2018bjl,Liu:2019krs,Florkowski:2021wvk,Liu:2021uhn,Fu:2021pok,Becattini:2021suc,Becattini:2021iol,Yi:2021ryh}. 
A consistent theoretical description
requires a theory of relativistic spin hydrodynamics, which extends standard hydrodynamics 
by a dynamical treatment of the spin tensor~\cite{Florkowski:2017ruc,Hattori:2019lfp,Speranza:2020ilk}. 

Theoretically, the polarization of hadrons is determined by the Pauli-Lubanski vector~\cite{Becattini:2020sww,Speranza:2020ilk,Tinti:2020gyh}. So far, calculations
of this quantity have assumed that the system \bbb{is in local equilibrium} at the point where hadrons decouple from 
the system~\cite{Becattini:2007sr,Becattini:2013vja,Becattini:2013fla,Becattini:2015ska,Becattini:2016gvu,Karpenko:2016jyx,Pang:2016igs,Xie:2017upb,Becattini:2021suc}. However, \bbb{local equilibrium can never be
reached, as it assumes that the Knudsen number Kn, which is the ratio of a typical microscopic length scale,
like the mean free path $\lambda_{\text{mfp}}$, to a typical macroscopic length scale, like the
hydrodynamic length scale $L_\text{hydro}$, vanishes. Therefore, in the hydrodynamic description of small systems, like heavy-ion collisions, the Knudsen number is never very small and dissipation always plays an important role~\cite{Heinz:2013th,Romatschke:2017ejr,Florkowski:2017olj}. Only if spin degrees of freedom evolve much faster than the bulk, we would expect them not to be influenced by dissipative effects. However,} results obtained from perturbative quantum 
chromodynamics~\cite{Kapusta:2019sad,Kapusta:2020npk} as well as in the framework of 
effective models~\cite{Kapusta:2019ktm,Ayala:2019iin,Ayala:2020ndx}
suggest that spin \bbb{dynamics} actually happens on the same or even a much larger time scale than that of
the collision as a whole (see also Refs.~\cite{Li:2019qkf,Yang:2020hri,Wang:2021qnt,Hongo:2022izs} for related work). 
\bbb{In this case, the Knudsen number for spin degrees of freedom is actually of the order of one, which would point towards a large influence of dissipation.} 

In order to understand (i) how the polarization of hadrons is influenced by dissipative effects, and (ii) \bbb{on which time scale spin degrees of freedom evolve}, one requires a theory of relativistic dissipative spin 
hydrodynamics. While a lot of effort has recently been invested towards establishing such a theory~\cite{Florkowski:2017ruc,Florkowski:2017dyn,Florkowski:2018myy,Florkowski:2018fap,Weickgenannt:2019dks,Bhadury:2020puc,Weickgenannt:2020aaf,Shi:2020htn,Speranza:2020ilk,Bhadury:2020cop,Singh:2020rht,Bhadury:2021oat,Peng:2021ago,Sheng:2021kfc,Sheng:2022ssd,Hu:2021pwh,Hu:2022lpi,Singh:2022ltu,Montenegro:2017rbu,Montenegro:2018bcf,Montenegro:2020paq,Gallegos:2021bzp,Hattori:2019lfp,Fukushima:2020ucl,Li:2020eon,She:2021lhe,Wang:2021ngp,Wang:2021wqq,Daher:2022xon,Gallegos:2020otk,Garbiso:2020puw,Cartwright:2021qpp,Hongo:2021ona,Weickgenannt:2022zxs,Weickgenannt:2022jes,Gallegos:2022jow,Bhadury:2022qxd,Cao:2022aku}, so far a general
derivation of relativistic second-order dissipative spin hydrodynamics from a microscopic theory is still missing.
In this Letter, we close this gap. We start from spin kinetic theory as derived from quantum field 
theory~\cite{Weickgenannt:2020aaf,Weickgenannt:2021cuo} 
(see Refs.~\cite{Florkowski:2018ahw,Weickgenannt:2019dks,Gao:2019znl,Hattori:2019ahi,Wang:2019moi,Liu:2020flb,Huang:2020kik,Wang:2020pej,Sheng:2021kfc,Wang:2022yli} for related work), and apply the method of 
moments~\cite{Denicol:2012es,Denicol:2012cn} appropriately generalized to
spin degrees of freedom. While a more detailed derivation is presented in Ref.~\cite{Weickgenannt:2022zxs},
here we merely outline the main steps and focus instead on answering 
questions (i) and (ii) listed above, which are important for heavy-ion phenomenology.

We use the following notation and conventions, $a\cdot b=a^\mu b_\mu$,
$a_{[\mu}b_{\nu]}\equiv a_\mu b_\nu-a_\nu b_\mu$, $a_{(\mu}b_{\nu)}\equiv a_\mu b_\nu+a_\nu b_\mu$, 
$g_{\mu \nu} = \mathrm{diag}(+,-,-,-)$, $\epsilon^{0123} = - \epsilon_{0123} = 1$. The dual of any rank-2 tensor 
$A^{\mu\nu}$ is defined as $\tilde{A}^{\mu\nu}\equiv \epsilon^{\mu\nu\alpha\beta}A_{\alpha\beta}$.
Denoting the fluid 4-velocity as $u^\mu$, we define the projector onto the subspace orthogonal to $u^\mu$ as
$\Delta^{\mu \nu} \equiv g^{\mu \nu} - u^\mu u^\nu$. The comoving derivative of a quantity $A$ is denoted as 
$\dot{A} \equiv u^\mu \partial_\mu A$, while the 3-space gradient is defined as $\nabla^\mu A \equiv \Delta^{\mu \nu}
\partial_\nu A$. The traceless, symmetric, and orthogonal projection of any rank-$\ell$ tensor is defined as
$A^{\langle\mu_1\cdots\mu_\ell\rangle}=\proj^{\mu_1\cdots\mu_\ell}_{\nu_1\cdots\nu_\ell} A^{\nu_1\cdots\nu_\ell}$,
where the projectors $\proj^{\mu_1\cdots\mu_\ell}_{\nu_1\cdots\nu_\ell}$ are
constructed from $\proj^{\mu\nu}$~\cite{DeGroot:1980dk}. In particular, for $\ell=2$ one obtains 
$\proj^{\mu\nu}_{\alpha\beta}\equiv (1/2)\proj^{(\mu}_\alpha \proj^{\nu)}_\beta-(1/3) \proj^{\mu\nu}\proj_{\alpha\beta}$.

\textit{Relativistic spin hydrodynamics ---}
The equations of motion of relativistic hydrodynamics read
\begin{align}
\partial_\lambda N^\lambda & = 0\;, &
\partial_\lambda T^{\lambda \mu} & = 0\;, &
\partial_\lambda S^{\lambda, \mu \nu} & = T^{[\nu \mu]}\;, \label{spin_cons}
\end{align}
where $N^\mu$ is the particle 4-current, $T^{\mu \nu}$ is the energy-momentum tensor, and
$S^{\lambda, \mu \nu}$ is the spin tensor.
The first two equations describe the conservation of particle number and energy-momentum.
If spin degrees of freedom are equilibrated, they only enter the description via the equation of state of the fluid.
However, if spin dynamics occurs on similar time and length scales as that of the fluid constituents, 
one requires additional evolution equations for the spin degrees of freedom. 
These are provided by the last equation in Eq.\ (\ref{spin_cons}), which follows from the conservation of 
total angular momentum. The system (\ref{spin_cons}) of equations of motion is then referred to as
relativistic spin hydrodynamics. Assuming an ideal fluid, the forms of $N^\mu$, $T^{\mu \nu}$, and $S^{\lambda, \mu \nu}$
are severely restricted and feature eleven unknowns: the inverse temperature $\beta_0 \equiv 1/T$, 
the ratio of chemical potential and temperature $\alpha_0 \equiv \mu/T$,
the fluid 4-velocity $u^\mu$ (which is time-like and normalized to one, $u^\mu u_\mu = 1$), and
the so-called spin potential $\Omega^{\mu \nu} = - \Omega^{\nu \mu}$. Thus, for ideal hydrodynamics, the above system 
of equations of motion is closed (once an equation of state for the fluid is provided) \cite{Florkowski:2017ruc}. 
In the general case,
however, additional equations for the dissipative quantities have to be specified. 

\textit{Boltzmann equation with nonlocal collision term---} We take as underlying microscopic theory the relativistic 
Boltzmann equation,
\begin{equation}
p\cdot \partial f=\mC[f]\;, \label{boltz}
\end{equation}
where ${\mC}[f]$ is the collision term. In standard kinetic theory at order $\mathcal{O}(\hbar^0)$, 
the range of interaction $\ell_\text{int}$
between particles is usually considered to be negligible compared to their mean free path $\lambda_\text{mfp}$, 
$\ell_\text{int} \ll \lambda_\text{mfp}$,
such that collisions are taken to occur at a single space-time point, and the collision term is local. 
However, at order $\mathcal{O}(\hbar)$, the collision term receives a nonlocal correction of order of the
Compton wavelength of the particles. For Dirac particles with spin 1/2 it was shown 
in Ref.~\cite{Weickgenannt:2020aaf} that, for binary elastic collisions, the collision term assumes the form
\begin{align}
{\mC}[f]   =& \int d\Gamma_1 d\Gamma_2 d\Gamma^\prime\,    {\mathcal{W}}\,  
[f(x+\Delta_1,p_1,\ms_1)f(x+\Delta_2,p_2,\ms_2)\n\\
&-f(x+\Delta,p,\ms)f(x+\Delta^\prime,p^\prime,\ms^\prime)]\; .
\label{finalcollisionterm}
\end{align}
Here, $f(x,p,\ms)$ is the spin-dependent single-particle distribution function in extended phase space, 
i.e., ordinary phase space extended by spin degrees of freedom. We also defined $d\Gamma\equiv dP\, dS(p)$ 
as the integration measure over on-shell momentum space, $dP\equiv d^4p\, \delta(p^2-m^2)$, and
spin space, $dS(p)\equiv  (\sqrt{p^2}/{\sqrt{3}\pi})  d^4\ms\, \delta(\ms\cdot\ms+3)\delta(p\cdot \ms)$.
The nonlocality of the collision term (\ref{finalcollisionterm}) manifests itself in the
fact that the distribution functions of the collision partners are taken at shifted space-time points, where
the space-time shift 
\begin{equation}
\label{deltanon}
\Delta^\mu\equiv -\frac{\hbar}{2m(p\cdot\hat{t}+m)}\, \epsilon^{\mu\nu\alpha\beta}p_\nu \hat{t}_\alpha \ms_{\beta}
\end{equation}
is of the order of the Compton wavelength of the particles.
Here, $\hat{t}^\mu \equiv (1,\boldsymbol{0})$ is a time-like unit vector defining the frame where $p^\mu$ is measured. 
Furthermore, $\mathcal{W}$ is the transition rate for the collision; for an explicit expression see Eq.\ (5) in
Ref.~\cite{Weickgenannt:2022zxs}. 

\textit{Generalized local equilibrium ---} 
Local equilibrium is usually defined by the condition that the collision term vanishes. 
Considering only the local part of the collision term, i.e., neglecting terms of order $\mathcal{O}(\Delta)$ in Eq.\
\eqref{finalcollisionterm}, this implies that (for Boltzmann statistics) the local-equilibrium distribution function 
must be an exponential function of a linear combination of quantities conserved in collisions, i.e.,
(expanded to first order in $\hbar$) it must be of the form
\begin{equation}
f_{\textrm{eq}}\equiv f_{0p}\left(1+\frac\hbar4 \Omega_{\mu\nu}\Sigma_\ms^{\mu\nu}\right)+\mathcal{O}(\hbar^2)\;, \label{feq}
\end{equation}
with $f_{0p}\equiv [{1}/{(2\pi\hbar)^3}] \exp({-\beta_0 E_p+\alpha_0})$, where $E_p \equiv u \cdot p$ and
$\Sigma_\ms^{\mu \nu} \equiv - \epsilon^{\mu \nu \alpha \beta} p_\alpha \ms_\beta /m$ is the dipole-moment tensor of
particles with momentum $p^\mu$ and spin $\ms^\mu$. Just like $\alpha_0$ and $\beta_0 u^\mu$
are Lagrange multipliers for the conservation of particle number and energy-momentum, respectively,
the spin potential $\Omega^{\mu \nu}$ is the Lagrange multiplier for the conservation of total angular momentum.

Expanding the collision term (\ref{finalcollisionterm}) in a Taylor series around the space-time point $x$, 
the local part of the collision term is the leading term of this Taylor series, while the
nonlocal part is the next-to-leading order term $\sim \Delta$, and because of Eq.\ (\ref{deltanon}) it is
also of order $\mathcal{O}(\hbar)$. As usual, the local part of the collision term
vanishes if we insert the local-equilibrium distribution function (\ref{feq}).
However, as shown in Ref.~\cite{Weickgenannt:2020aaf}, to first order in an expansion in powers of $\hbar$,
the sum of the nonlocal collision term and the order $\mathcal{O}(\hbar)$ contribution to the local collision term 
rigorously vanishes only in \textit{global\/}
equilibrium, i.e., we have to demand in addition that $\alpha_0 = \textit{const.}$ and $\beta_0 u_\mu$ is
a Killing vector, $\partial_{(\mu} (\beta_0 u_{\nu)}) = 0$. In addition, the spin potential equals the so-called thermal
vorticity, $\Omega_{\mu \nu} =  \varpi_{\mu \nu} \equiv - \frac 1 2 \partial_{[\mu} (\beta_0 u_{\nu]})= \textit{const.}$, 
which follows from the conservation of total angular momentum in binary elastic collisions.

The local-equilibrium distribution function (\ref{feq}) is not a solution of the Boltzmann equation (\ref{boltz}).
However, hydrodynamics can be considered to be 
an effective theory for the long-wavelength, small-frequency limit of the latter, which implies a clear separation
of microscopic and macroscopic length scales. If the typical microscopic length scale is taken as
the mean free path $\lambda_\text{mfp}$ of the particles between collisions, and the macroscopical
length is identified as the typical scale $L_\text{hydro}$ over which the hydrodynamic fields
$\alpha_0$, $\beta_0$, and $u^\mu$ vary, then
the validity of hydrodynamics is ensured by the condition $\lambda_\text{mfp} \ll L_\text{hydro}$, or equivalently,
that the Knudsen number $\text{Kn} \equiv \lambda_\text{mfp}/ L_\text{hydro} \ll 1$.
In this limit, the gradient terms on the left-hand side of the Boltzmann equation for $f_{\textrm{eq}}$
are of the order of the inverse hydrodynamical length scale, i.e., 
\begin{equation}
\beta_0 \partial_\mu \alpha_0  \,, \;
\partial_{(\mu} (\beta_0 u_{\nu)}) \, , \; 
\beta_0 \partial_\mu \Omega^{\alpha \beta}  \sim \mathcal{O}(\beta_0/L_\text{hydro})\;.
\label{localeq_cond3}
\end{equation}
Consequently, inserting the local-equilibrium form (\ref{feq}) into the nonlocal collision term, we observe that
the sum of the order $\mathcal{O}(\hbar)$ contribution to the local collision term 
and the nonlocal collision term is of order 
\begin{align}
\lefteqn{\frac{1}{f_{0p}} \left( \frac{\hbar}{4} \Sigma_\ms^{\mu \nu} \Omega_{\mu \nu} 
+ \Delta^\mu \partial_\mu\right) f_{0p}}
\nonumber \\
&=  \frac{1}{2} \Delta^{[\mu} p^{\nu]} (\varpi_{\mu \nu} -\Omega_{\mu \nu}) 
+\mathcal{O}\left(\Delta/ L_\text{hydro} \right)\;, \label{hbar_coll_power}
\end{align}
where we used Eq.\ (\ref{localeq_cond3}) as well as the conservation of
total angular momentum, $J^{\mu\nu}\equiv \Delta^{[\mu}p^{\nu]}+(\hbar/2)\Sigma_\ms^{\mu\nu}$.

Since $\Delta \sim \hbar/m$ is at most of the order of the range of the interaction and 
thus much smaller than the mean free path, we have the following ordering of scales:
\begin{equation} \label{ordering_scales}
\Delta \lesssim \ell_\text{int} \ll \lambda_\text{mfp}\ll L_\text{hydro}\;.
\end{equation}
Therefore, the $\mathcal{O}\left(\Delta/ L_\text{hydro} \right)$ contribution to the collision term in Eq.\
\eqref{hbar_coll_power} is much smaller than the
Knudsen number $\text{Kn} \equiv \lambda_\text{mfp}/L_\text{hydro} \ll 1$. 
This leads us to an \textit{extension of the definition of local equilibrium:} instead of demanding that
the collision term vanishes exactly, we only demand that it vanishes \textit{up to terms of order\/} $\Delta / L_\text{hydro}$.
In order for the whole collision term to fulfill this condition, the first term in the last line of 
Eq.\ \eqref{hbar_coll_power} must also be of order $\mathcal{O}\left(\Delta/ L_\text{hydro} \right)$.
This is fulfilled if 
\begin{equation} \label{spin-pot_cond}
\Omega_{\mu \nu} = \varpi_{\mu \nu} + \mathcal{O}(\beta_0 /L_\text{hydro})\;.
\end{equation}
This condition is natural: by definition, local equilibrium constitutes a correction of order 
$\mathcal{O}(\beta_0/ L_\text{hydro})$ to global equilibrium, cf.\ Eq.\ (\ref{localeq_cond3}), and
$\Omega_{\mu \nu} = \varpi_{\mu \nu}$ in global equilibrium.

\textit{Modified power counting ---} 
The conditions of global equilibrium do not restrict the value of the thermal vorticity $\varpi_{\mu \nu}$. 
Assuming that gradients of inverse temperature, $\nabla^\alpha \beta_0$, and fluid acceleration, $\dot{u}^\alpha$,
are of order $\mathcal{O}(1/L_\text{hydro})$, this is equivalent to the statement that the fluid
vorticity $\omega^{\alpha \beta} \equiv ({1}/{2}) \nabla^{[\alpha} u^{\beta]}$ is not constrained. 
Let us introduce the length scale $\ell_\text{vort}$ which characterizes the magnitude of the fluid 
vorticity, $\omega^{\alpha \beta} \sim \mathcal{O}(1/\ell_\text{vort})$. 
In order for the expansion in powers of $\hbar$ to remain meaningful, we have to assume that
$\hbar \Omega_{\mu \nu} \Sigma_\ms^{\mu \nu} \ll 1$, cf.\ Eq.\ \eqref{feq}. Taking
$\hbar/m \sim \Delta$ and the typical momenta $p^\mu \sim 1/\beta_0$, we
then arrive at the condition that $\Delta /\ell_\text{vort} \ll 1$. In principle, this ratio introduces a new
small parameter for power counting. However, in order to keep the discussion as simple as possible, in the
following we will assume that $\Delta /\ell_\text{vort} \sim \text{Kn}$. Using Eq.\ \eqref{ordering_scales},
$\ell_\text{vort} \sim (\Delta/\lambda_\text{mfp}) L_\text{hydro} \ll L_\text{hydro}$, i.e., it
can be much smaller than the scale associated with other hydrodynamical gradients.
\bbb{This means that rapidly rotating systems sufficiently close to local equilibrium, such as the quark-gluon plasma in noncentral heavy-ion collisions or rotating neutron stars, can be described by a hydrodynamic theory with large values for the fluid vorticity.}

\textit{Moment expansion ---} The standard method of moments is based on an expansion 
of the single-particle distribution function in terms of
irreducible moments of the deviation of the latter from a reference distribution, usually taken to be the
one in local thermodynamic equilibrium \cite{Denicol:2012cn}. In phase space extended by
spin degrees of freedom, this expansion needs to be generalized to also include moments of the spin vector $\ms^\mu$.
Introducing
$\langle \cdots \rangle  \equiv \int d\Gamma\, (\cdots) f,$
$\langle \cdots \rangle_{\textrm{eq}}  \equiv \int d\Gamma\, (\cdots) f_{\textrm{eq}},$
and $\langle \cdots \rangle_\delta  \equiv \langle \cdots \rangle - \langle \cdots \rangle_{\textrm{eq}}$,
the spin moments are defined as~\cite{Weickgenannt:2022zxs}
\begin{equation}
 \tau_n^{\mu,\mu_1\cdots \mu_l}\equiv \langle E_p^n\, \ms^\mu p^{\langle \mu_1}\cdots p^{\mu_l\rangle}\rangle_\delta\; . 
 \label{spinmom}
\end{equation}

The distribution function can now be expanded in terms of the moments  
\eqref{spinmom} as~\cite{Weickgenannt:2022zxs}
\begin{align}
& f(x,p,\ms) =f^{(0)}(x,p)+f_{0p}\Bigg[ \frac\hbar4\Omega_{\mu\nu}\Sigma_\ms^{\mu\nu}
-\sum_{l=0}^\infty \sum_{n\in \mathbb{S}_l}\mathcal{H}_{pn}^{(l)}\n\\
 &\times\left(g_{\mu\nu}-\frac{p_{\langle\mu\rangle}}{E_p} u_\nu \right)
 \ms^\nu \tau_n^{\lmur,\mu_1\cdots\mu_l}p_{\langle\mu_1}\cdots p_{\mu_l\rangle}\Bigg]\;, \label{distrnoneqfin}
 \end{align} 
where the spin-independent part $f^{(0)}(x,p)$ and the coefficients $ \mathcal{H}_{pn}^{(l)}$ are given in 
Ref.~\cite{Denicol:2012cn}. 

\textit{Decomposition of the spin tensor ---} 
Our derivation of dissipative spin hydrodynamics for Dirac particles is based on the spin tensor in the 
Hilgevoord--Wouthuysen (HW) 
pseudo-gauge~\cite{Weickgenannt:2020aaf,Speranza:2020ilk}
\begin{align}
S^{\lambda,\mu\nu}&= \frac12\left\langle p^\lambda  \Sigma_{\ms}^{\mu\nu}\right\rangle 
-\frac{\hbar}{4m^2}\partial^{[\nu} \left\langle p^{\mu]} p^\lambda \right\rangle\;. \label{conscurr}
\end{align}
Inserting Eq.\ (\ref{distrnoneqfin}) into \eq\eqref{conscurr} we find
\begin{align}
   S^{\lambda,\mu\nu}={}&u^\lambda \tilde{\mathfrak{N}}^{\mu\nu}+\Delta^\lambda_{\, \alpha}  
   \tilde{\mathfrak{P}}^{\alpha\mu\nu}+ u_{(\alpha}\tilde{\mathfrak{H}}^{\lambda)\mu\nu\alpha}
   +\tilde{\mathfrak{Q}}^{\lambda\mu\nu}\n\\
    &+\frac{\hbar}{2m}\partial^{[\nu}\left[\epsilon u^{\mu]} u^\lambda-\Delta^{\mu]\lambda}(P_0+\Pi)
    +\pi^{\mu]\lambda}\right]\; , \label{spitendeco}
\end{align}
with the energy density $\epsilon\equiv \langle E_p^2\rangle$, the thermodynamic pressure $P_0$, 
the bulk viscous pressure $\Pi\equiv -(1/3)\langle \proj^{\mu\nu} p_\mu p_\nu\rangle -P_0$, and the shear-stress tensor 
$\pi^{\mu\nu}\equiv \langle p^{\langle\mu} p^{\nu\rangle}\rangle$. The quantities in the first line of \eq\eqref{spitendeco} 
are given by the duals of the spin-energy, spin-pressure, spin-diffusion, and spin-stress tensors, respectively,
\begin{eqnarray}
 {\mathfrak{N}}^{\mu\nu}&\equiv& -\frac{1}{2m}u^{\mu} \langle E_p^2\, \ms^{\nu}\rangle_{\textrm{eq}} 
 -\frac{1}{2m}u^\mu \mn^\nu ,\n\\
 {\mathfrak{P}}^{\mu}&\equiv& -\frac{1}{6m} \langle \Delta^{\rho\sigma} p_\rho p_\sigma\, \ms^\mu\rangle_{\textrm{eq}} 
 -\frac{1}{6m}\left(m^2 \map^\mu-\mn^\mu\right),\n\\
{ \mathfrak{H}}^{\lambda\mu}&\equiv& -\frac{1}{2m}\langle  E_p p^{\langle\lambda\rangle}  \ms^{\mu}\rangle_{\textrm{eq}} 
- \frac{1}{2m} \mathfrak{h}^{\lambda\mu},\n\\
 {\mathfrak{Q}}^{\lambda\mu\nu}&\equiv& -\frac{1}{2m}\mathfrak{q}^{\lambda\mu\nu}\;,
\end{eqnarray}
with 
the nonequilibrium spin moments
\begin{align}
\mn^\nu&\equiv\taum_2^\nu\;, & \map^\mu &\equiv \tau_0^\mu\;, 
&\mathfrak{h}^{\lambda\mu}&\equiv\taum_1^{\mu,\lambda}\;, 
& \mathfrak{q}^{\lambda\mu\nu}\equiv \tau^{\lambda,\mu\nu}_{0} \; . \label{nphqdef}
\end{align}
Note that the components of all spin moments which are parallel to $u^\mu$ 
in the first index can be expressed as a linear combination of the orthogonal components, see
Eq.\ (68) of Ref.\ \cite{Weickgenannt:2022zxs}. 
After accounting for this, the moments (\ref{nphqdef}) contain 24 degrees of freedom, but, as shown in 
Ref.~\cite{Weickgenannt:2022zxs}, 
a certain combination of these spin moments vanishes after imposing Landau matching conditions.
The remaining 18 independent components of the spin tensor are then
\begin{align}
  \map^\lmur &\equiv \tau_0^\lmur\;, & \mfv^{\mu\nu}&\equiv\taum_1^{(\lmur,\nu)}\;, 
  & \mathfrak{q}^{\langle\lambda\rangle\mu\nu}\equiv \tau^{\langle\lambda\rangle,\mu\nu}_{0} \; . \label{dynspin}
\end{align}
In our framework, these dissipative spin moments are treated as dynamical variables.

\textit{Equations of motion for the spin moments ---}
The equations of motion for the spin moments \eqref{spinmom}
can be derived from the Boltzmann equation \eqref{boltz} without any further approximation, see 
Ref.~\cite{Weickgenannt:2022zxs} for details. Defining the collision integrals
\begin{equation}
\mC_r^{\mu,\langle \mu_1\cdots \mu_n\rangle}\equiv \int d\Gamma\, 
E_p^r\, p^{\langle\mu_1} \cdots p^{\mu_n\rangle} \ms^\mu \mC[f]\;, \label{coli}
\end{equation}
one finally obtains relaxation-type equations for the spin moments of the form
\begin{equation}
    \dot{\tau}_r^{\langle\mu\rangle,\langle\mu_1\cdots\mu_l\rangle}-\mC_{r-1}^{\lmur,\langle \mu_1\cdots \mu_n\rangle}
    =\mathcal{O}(\Omega\partial) + \mathcal{O}(\partial^2)\;, \label{taudot}
\end{equation}
with $\mathcal{O}(\Omega \partial)$ denoting terms of first order in $\Omega^{\mu \nu}$ and gradients and 
$\mathcal{O}(\partial^2)$ denoting terms of first order in the product of gradients and dissipative quantities.
The terms of order $\mathcal{O}(\Omega \partial)$ arise from the $\mathcal{O}(\hbar)$ contribution to the local-equilibrium
distribution function. These terms constitute part of the Navier-Stokes terms for the
spin moments. The other part arises from the $\mathcal{O}(\hbar)$ part of the collision term. However, assuming that 
the spin potential is small, the latter terms will give the leading-order contribution on long time scales. 
Thus, we will focus on their effects in the following. 

\textit{Collision terms --- }
In order to evaluate the collision term, we assume that the transition rate $\mathcal{W}$ 
does not depend on the phase-space spin variables. However, this does not mean that particle spins cannot change 
during a collision, since we still have nonlocal contributions to the collision term which mutually 
convert orbital angular momentum and spin. 
Consequently, inserting \eq\eqref{finalcollisionterm} with \eq\eqref{distrnoneqfin}
into \eq\eqref{coli} and using the conservation of total angular momentum, the linearized collision term reads
\begin{align}
&{\mC}_{r-1}^{\mu,\langle\mu_1\cdots\mu_n\rangle}
=-\sum_{n\in\mathbb{S}_l} B_{rn}^{(l)} \taum_n^{\mu,\langle\mu_1\cdots\nu_l\rangle} \!\!+ \!\!
\int [d\Gamma]\, \mathcal{W}\, E_p^{r-1} f_{0p} f_{0p^\prime} \n\\
&\times p^{\langle\mu_1}\cdots p^{\mu_n\rangle} \ms^\mu  \left[  \frac\hbar4 (\varpi_{\alpha\beta}-\Omega_{\alpha\beta}) 
\Sigma_\ms^{\alpha\beta}+\frac12 \partial_{(\beta}\beta_{\alpha)} \Delta^\beta p^\alpha \right]\, , \label{waisdjd}
\end{align}
where $[d \Gamma] \equiv [dP] [dS]$, with $[dP] \equiv dP dP' dP_1 dP_2$ and similarly for $[dS]$.
Furthermore, $\mathbb{S}_l$ is the set of indices of those spin moments that are treated dynamically.
The scalar collision integrals are defined as
\begin{align}
B_{rn}^{(l)} \equiv& -\frac{16}{2l+1} \proj^{\nu_1\cdots\nu_l}_{\mu_1\cdots\mu_l}  \int [dP] \mathcal{W}_0
f_{0p} f_{0p^\prime} E_p^{r-1}\n\\ &\times p^{\langle\mu_1}\cdots p^{\mu_l\rangle}  
\mathcal{H}_{pn}^{(l)}p_{\langle\nu_1}\cdots p_{\nu_l\rangle}\;, \label{BDelB}
\end{align}
where $\mathcal{W}_0 \equiv (1/2)^4 \int [dS] \mathcal{W}$.
Performing the integrations in \eq\eqref{waisdjd} and choosing $\hat{t}^\mu=u^\mu$ , 
see Ref.~\cite{Weickgenannt:2022zxs} for details, we find for the three lowest tensor-rank collision terms
\begin{align}
\mC_{r-1}^\mu&=-\sum_{n\in\mathbb{S}_0} B_{rn}^{(0)} \taum_n^{\mu}+g^{(0)}_{r} \left(\tilde{\Omega}^{\mu\nu}
-\tilde{\varpi}^{\mu\nu}\right) u_\nu\;,\n\\
\mC_{r-1}^{(\mu,\nu)}&=-\sum_{n\in\mathbb{S}_1} B_{rn}^{(1)} \taum_n^{(\mu,\langle\nu\rangle)}\;,\n\\
\mC_{r-1}^{\mu,\nu\lambda}&=  -\sum_{n\in\mathbb{S}_2} B_{rn}^{(2)} \taum_n^{\mu,\langle\nu\lambda\rangle}
+ h_r^{(2)} \beta_0 \sigma_\rho^{\ \langle\nu} \epsilon^{\lambda\rangle\mu\alpha\rho} u_\alpha\;, \label{ccbaridontknow}
\end{align}
with $\sigma^{\mu\nu}\equiv\partial^{\langle\mu}u^{\nu\rangle}$ being the shear tensor and
\begin{align}
g_r^{(0)} \equiv & \frac{4\hbar}{m}\int [dP]\, \mathcal{W}_0 E_p^{r} f_{0p} f_{0p^\prime}\;, \n\\
h_r^{(2)} \equiv & -\frac{16\hbar}{15m}\int [dP]\, \frac{\mathcal{W}_0 }{E_p+m} E_p^{r-1} f_{0p} f_{0p^\prime} 
(\proj^{\alpha\beta} p_\alpha p_\beta)^2\;.
\end{align}
The first term on the right-hand sides of Eqs.~\eqref{ccbaridontknow} is a linear combination of all spin moments of the 
same tensor rank, respectively, which originates from the local collision term. Inverting the 
coefficient matrices $B_{rn}^{(l)}$ one obtains the relaxation times for the spin moments.
On the other hand, the remaining terms in \eqs\eqref{ccbaridontknow} appear due the nonlocality of the collision term 
and do not depend on the spin moments themselves, but on gradients of fluid velocity and temperature, or the difference 
between the spin potential and the thermal vorticity. While in global equilibrium all these terms vanish, they will contribute 
to the Navier-Stokes limit of the components of the spin tensor.

\textit{14+24-moment approximation ---} 
For a viable theory of spin hydrodynamics, the infinite system (\ref{taudot}) of 
equations of motion for the spin moments has to be truncated and closed. The lowest possible truncation involves
the 24 components of the spin tensor listed in Eqs.~\eqref{nphqdef}. All other spin moments can be expressed 
in terms of these via a linear relation, see Eq.\ (107) of Ref.~\cite{Weickgenannt:2022zxs}.
We refer to this truncation as the 14+24-moment approximation, since it adds the 24 degrees of freedom of the spin tensor 
to the 14 dynamical moments of Israel-Stewart theory. 
Introducing $\mathfrak{T}^{(l)}\equiv \left( B^{(l)}\right)^{-1}$, multiplying Eqs.\ 
\eqref{taudot} for the moments of interest with $\mathfrak{T}_{nr}^{(l)}$ and summing over $r$ 
with $\mathbb{S}_0=\{0\}$, $\mathbb{S}_1=\{1\}$, and $\mathbb{S}_0=\{0\}$ (corresponding to the
indices pertaining to the dynamically treated moments \eqref{dynspin}), the resulting 
equations of motion for the moments \eqref{dynspin} are
\begin{align}
& \tau_{\mft}  \proj^\mu_\nu \frac{d}{d\tau}  {\mft}^{\langle\nu\rangle}+\mft^\lmur=
 \mathfrak{e}\left(\tilde{\Omega}^{\mu\nu}-\tilde{\varpi}^{\mu\nu}\right) u_\nu
 +\mathcal{O}(\Omega\partial,\partial^2)\; ,\n\\
& \tau_{\mfv}\proj^\mu_\lambda \proj^\nu_\rho \frac{d}{d\tau}{\mfv}^{\lambda\rho}+\mfv^{\mu\nu} =
 \mathcal{O}(\Omega\partial,\partial^2)\, , \n\\
&\tau_\mfw\proj^\mu_\rho \proj^{\nu\lambda}_{\alpha\beta}\frac{d}{d\tau}{\mfw}^{\langle\rho\rangle\alpha\beta}
+\mfw^{\lmur\nu\lambda}= - \mathfrak{d} \beta_0 \sigma_\rho^{\ \langle\nu} 
\epsilon^{\lambda\rangle\mu\alpha\rho} u_\alpha\n\\
&+\mathcal{O}(\Omega\partial,\partial^2)\, , \label{spreeq}
 \end{align}
where the spin relaxation times are given by
\begin{align}
\tau_{\mft}&= \mathfrak{T}_{00}^{(0)}\;,&
\tau_\mfv&=   \mathfrak{T}_{11}^{(1)}\;,&
\tau_\mfw&= \mathfrak{T}_{00}^{(2)}\;, \label{srt}
\end{align}
and we furthermore defined
$\mathfrak{e}= \tau_\mft g_0^{(0)}$ and $\mathfrak{d}= \tau_\mfw h_r^{(2)}$. \bbb{While the full second-order equations of motion are shown in Ref.~\cite{Weickgenannt:2022zxs}, for the following discussion only the terms given in \eqs\eqref{spreeq} are relevant.
One can see from these equations that} the dynamical spin moments relax on the time scales \eqref{srt} to the Navier-Stokes terms
on the right-hand sides of \eqs\eqref{spreeq}. Note that on account of Eq.~\eqref{spin-pot_cond}, these terms
are of the same order as the Navier-Stokes terms for the standard dissipative quantities (up to a factor
of $\hbar$ contained in the functions $\mathfrak{e}^{(0)}$ and $\mathfrak{d}$, which arises from the
nonlocal collision term).
At this point, we remark that also the terms $\sim O(\Omega \partial)$ 
contribute to the Navier-Stokes limit, as the spin potential is an equilibrium 
quantity. Hence the Navier-Stokes term for ${\mfv}^{\lambda\rho}$ is in general nonvanishing, although for small spin 
potential it is smaller than those of the other spin moments. 

\begin{figure}[t]
        \includegraphics[width=0.45\textwidth]{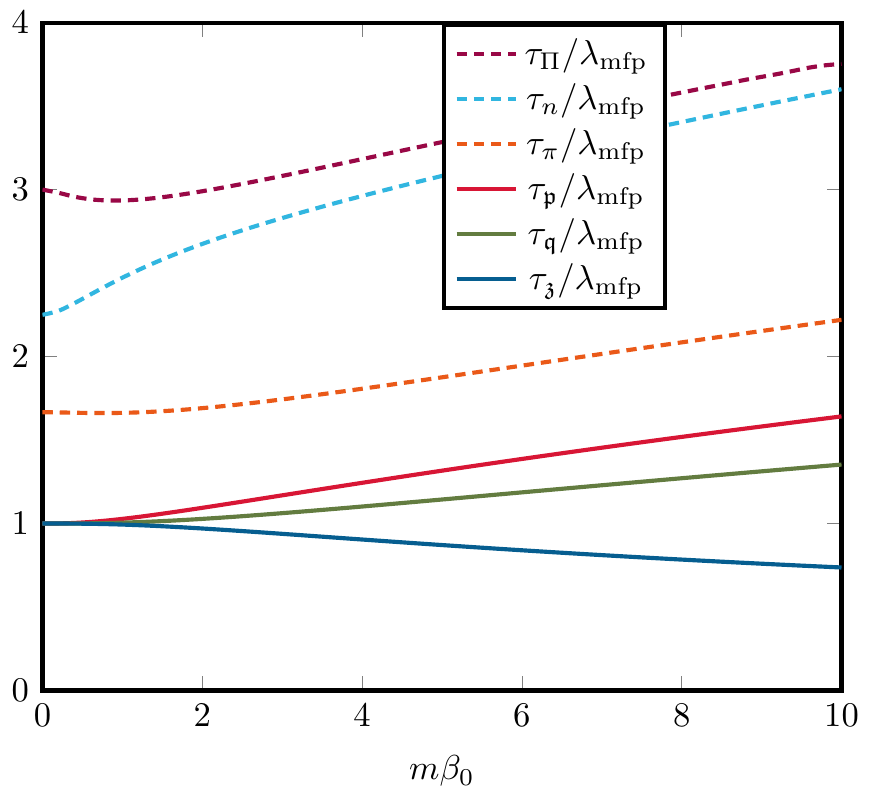}
		\caption{Relaxation times for 
		$\mathfrak{p}^\mu$,  $\mathfrak{z}^{\mu\nu}$, $\mathfrak{q}^{\lambda\mu\nu}$
		(full lines), in comparison to those for the 
		standard dissipative quantities $\Pi$, $n^\mu$, $\pi^{\mu\nu}$ (dashed lines), computed within
		the standard 14-moment approximation~\cite{Denicol:2012cn},
		as a function of $m\beta_0$~\cite{Weickgenannt:2022zxs}.} 
		\label{fig:tau_spin}
	\end{figure}

\textit{Spin relaxation times ---}
For a constant cross section, \eqs\eqref{srt} can be straightforwardly computed, 
see Ref.~\cite{Weickgenannt:2022zxs} for details. The resulting spin relaxation times are shown in Fig.~\ref{fig:tau_spin}
in units of the mean free path $\lambda_\text{mfp}$ for different values of mass over temperature. 
We find that they are of the same order of magnitude, but slightly smaller than those for the usual dissipative currents. \bbb{It should be noted that in our framework the condition (\ref{ordering_scales}) forms a lower bound on the
mass. Thus, the latter cannot be taken to be arbitrarily small, and $m\beta_0\rightarrow0$ actually corresponds to the limit $T\rightarrow\infty$, and not $m\rightarrow 0$. The convergence of the spin relaxation times in this limit is most likely an artifact of our assumption of a constant cross section.}


\textit{Pauli-Lubanski vector ---}
In the context of heavy-ion collisions, the observable polarization, the so-called Pauli-Lubanski vector, is given 
by \cite{Becattini:2020sww,Speranza:2020ilk}
\begin{equation} \label{Pauli}
\Pi^\mu(p) = \frac{1}{2\mathcal{N}} \int d\Sigma \cdot p \, dS(p)\, \ms^\mu f(x,p,\ms)\;,
\end{equation}
with $d\Sigma_\mu$ denoting the integration over the freeze-out hypersurface and
$\mathcal{N}\equiv  \int d\Sigma \cdot p \, dS(p) f(x,p,\ms)$.
In order to obtain the leading-order contribution in the Navier-Stokes limit, we insert \eq\eqref{distrnoneqfin}
into Eq.~\eqref{Pauli} and replace all non-dynamical spin moments by the dynamical ones 
$\mft^\lmur$, $\mfv^{\mu\nu}$, and $\mfw^{\lmur\nu\lambda}$. For the latter, we then insert the right-hand sides of 
\eqs\eqref{spreeq}. Thus we find up to order $\mathcal{O}(\Omega\partial,\partial^2)$
\begin{align}
&\Pi_\text{NS}^\mu(p) =   
\int d\Sigma\cdot p \, \frac{f_{0p}}{2\mathcal{N}} \bigg\{ -\frac{\hbar}{2m} \tilde{\Omega}^{\mu\nu}p_\nu
+ \left(g^\mu_\nu- \frac{ u^\mu p_{\langle\nu\rangle}}{E_p}\right)\n\\ &\times\left[\mathfrak{e}\chi_\mft\left(\tilde{\Omega}^{\nu\rho}-\tilde{\varpi}^{\nu\rho}\right) u_\rho-\chi_\mfw \mathfrak{d} \beta_0 \sigma_\rho^{\ \langle\alpha} 
\epsilon^{\beta\rangle\nu\sigma\rho} u_\sigma p_{\langle\alpha} p_{\beta\rangle}\right]\bigg\} 
 \;,\label{pollns}
\end{align}
with $\chi_\mft$ and $\chi_\mfw$ being functions of $E_p$, $\alpha_0$, and 
$\beta_0$~\cite{Weickgenannt:2022zxs}. We see from \eq\eqref{pollns} that, through the nonlocal collision term, 
the Pauli-Lubanski vector obtains contributions from the 
difference between the spin potential and thermal vorticity, as well as from the shear tensor. 
Terms proportional to the shear tensor have recently been found to be important for the experimentally measured 
Lambda polarization in heavy-ion collisions~\cite{Liu:2021uhn,Fu:2021pok,Becattini:2021suc,Becattini:2021iol,Liu:2021nyg,Florkowski:2021xvy}
\bbb{(see also Refs.~\cite{Hidaka:2017auj,Fang:2022ttm} for related work in the massless case). Although the form of this term in our approach is similar to the one obtained in Refs.~\cite{Liu:2021uhn,Fu:2021pok,Becattini:2021suc,Becattini:2021iol}, it may have a different origin, since in our framework it emerges from the nonlocal collision term. 
We note that a term $\sim \tilde{\Omega}^{\nu\rho}-\tilde{\varpi}^{\nu\rho}$ was also found in Ref.\ \cite{Buzzegoli:2021wlg}. We furthermore remark that, although the form of the Pauli-Lubanski vector \eqref{Pauli} is independent of the pseudo-gauge, the pseudo-gauge choice implicitly enters \eq\eqref{pollns} through the truncation of the moment expansion \eqref{spinmom}, for which only the components of the HW spin tensor are considered. If we had chosen a different pseudo-gauge, also our truncation of the moment expansion would be different.}

\textit{Conclusions ---}
In this Letter, we derived the equations of motion of dissipative spin hydrodynamics~\cite{Weickgenannt:2022zxs},
starting from the Boltzmann equation with a nonlocal collision term, which  
required us to extend the concept of local thermodynamical equilibrium and to devise a 
new power-counting scheme. 
We then applied the method of moments 
in the 14+24-moment approximation to derive relaxation-type equations for the dissipative spin moments.
Interestingly, the relaxation time scales emerge from the local part of the collision term 
in the Boltzmann equation, while the nonlocal part of the latter contributes to the Navier-Stokes values 
for the spin moments. 

We found that the spin relaxation times are of the same order of magnitude as those of the usual 
dissipative currents, which answers question (ii) posed in the Introduction. \bbb{Qualitatively, we may compare this result to} the findings 
of Refs.~\cite{Kapusta:2019sad,Kapusta:2019ktm,Kapusta:2020npk}, which
studied helicity-changing processes to estimate the time scale for spin relaxation, \bbb{with the result that these processes lead to comparably large spin relaxation times.}
However, these are not necessarily the only processes which lead to spin \bbb{relaxation}. 
Namely, spin \bbb{dynamics is also influenced by} collisions which only exchange momentum,
but not spin, since the polarization of a fluid element can also change
through the mere transport of particles, carrying the polarization to a different fluid element. 
This effect provides the main contribution to spin diffusion and naturally leads to spin relaxation times comparable to 
those of other dissipative currents equilibrating through the standard (spin-independent) collision term. 

The nonlocal collision term is responsible for aligning the spin with the vorticity~\cite{Weickgenannt:2020aaf}.
As long as global equilibrium is not reached, the nonlocal
collision term contributes to the Navier-Stokes values of the spin moments, with terms proportional to the difference 
between the spin potential and the thermal vorticity, and proportional to the shear tensor. 
The precise way in which dissipative effects enter the polarization
answers question (i) posed in the Introduction. 
Since the time scale of a heavy-ion collision is too short to reach global equilibrium, the polarization of
particles in such collisions cannot only be determined by the thermal vorticity. 
Finally, it was found that thermal shear may play an important role in the description of 
Lambda polarization~\cite{Liu:2021uhn,Fu:2021pok,Becattini:2021suc,Becattini:2021iol,Liu:2021nyg,Florkowski:2021xvy}. 
The findings of this Letter confirm the importance of such terms and provide an interpretation of their origin 
in the context of kinetic theory. 

\textit{Acknowledgements ---}
The authors thank U.\ Heinz and J.\ Noronha for enlightening discussions. The work of D.H.R., D.W., and N.W.\ is supported by the
Deutsche Forschungsgemeinschaft (DFG, German Research Foundation)
through the Collaborative Research Center Trans\-Regio
CRC-TR 211 ``Strong-interaction matter under extreme conditions'' -- project number
315477589 -- TRR 211 and by the State of Hesse within the Research Cluster
ELEMENTS (Project ID 500/10.006).
D.W.\ acknowledges support by the Studienstiftung des deutschen Volkes 
(German Academic Scholarship Foundation).

\bibliography{biblio_paper_long}{}

\end{document}